\documentclass[aps,prl,10pt,twocolumn,groupedaddress]{revtex4-1}

\usepackage[pdftex]{graphicx}      
\usepackage{amsmath}

\def\UoB{
	School of Physics and Astronomy, 
	University of Birmingham, 
	Edgbaston, Birmingham B15 2TT, 
	United Kingdom}

\begin{document}

\title{Directional bistability and nonreciprocal lasing with cold atoms in a ring cavity}

\author{B. Megyeri}
\author{G. Harvie}
\author{A. Lampis}
\author{J. Goldwin}
\email[Corresponding author:~]{j.m.goldwin@bham.ac.uk}
\affiliation{\UoB}


\begin{abstract}
We demonstrate lasing into counter-propagating modes of a ring cavity using a gas of cold atoms as a gain medium. The laser operates under the usual conditions of magneto-optical trapping with no additional fields. We characterize the threshold behavior of the laser and measure the second-order optical coherence. The laser emission exhibits directional bistability, switching randomly between clockwise and counter-clockwise modes, and a tuneable nonreciprocity is observed as the atoms are displaced along the cavity axis.
\end{abstract}

\maketitle

The unique features of cold atomic gases, including coherences in electronic, spin, and motional degrees of freedom, have enabled the realization of lasers with a wide range of unconventional characteristics. Extremely narrow gain resonances have been engineered, leading to sub-luminal group velocities, extended coherence times, and reduced sensitivity to environmental perturbations~\cite{Hilico92, Guerin08, Vrijsen11, Bohnet12}. Cold-atom lasers operating on dipole-forbidden transitions~\cite{Norcia16,Gothe17} have shown promise as active optical frequency standards~\cite{Norcia18}. Both spatial order ~\cite{Schilke12} and disorder~\cite{Baudouin13} have been shown to sustain mirrorless lasing, and momentum-space coherences have enabled collective atomic recoil lasing~\cite{Kruse03}. The quantum limit of lasing has been realized with a single atom, resulting in nonclassical photon statistics in the emitted field~\cite{McKeever03}.

Missing from this body of work is a bidirectional ring laser emitting light into counter-propagating traveling-wave cavity modes. Such lasers can exhibit rich inter-mode dynamics and nonreciprocal effects, breaking the symmetry between propagation directions~\cite{Kravtsov99}. A prominent example of a nonreciprocal laser is the ring laser gyro, which exploits the Sagnac effect to convert angular rotation into a beat frequency between clockwise (CW) and counterclockwise (CCW) modes~\cite{SargentScullyLamb,Stedman97}. A nonreciprocal dye ring laser was used to investigate stochastic resonances~\cite{McNamara88} and semiconductor microring lasers have been used for studies of parity-time symmetry breaking~\cite{Peng14, Feng14, Hodaie14} and chiral effects~\cite{Redding12, Sarma15, Gustave15}. 

In this Letter, we demonstrate bidirectional lasing with cold atoms in a ring cavity. The laser is pumped by the same fields used for magneto-optical cooling and trapping, with no additions or modifications. The onset of lasing is evidenced through threshold behavior with increasing atom number and a transition from photon bunching to second-order coherence. The laser exhibits bistable switching between clockwise and counter-clockwise modes, and nonreciprocal behavior is observed.

\begin{figure}[ht]
\centering
\includegraphics[width=0.6\columnwidth]{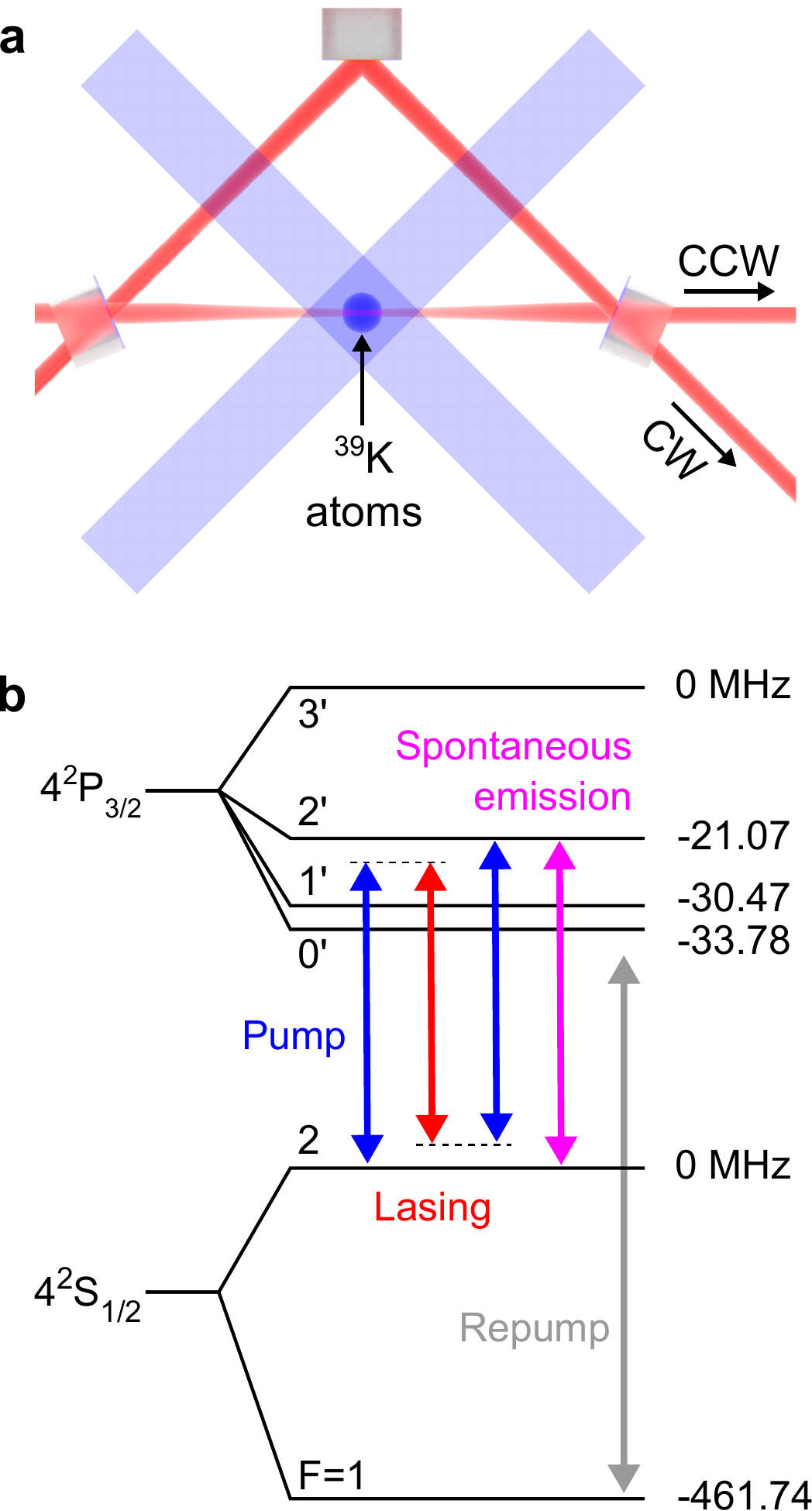}
\caption{Bidirectional cold atom ring laser. (a) Experiment schematic. A cloud of potassium-39 atoms is cooled and pumped (blue beams), causing laser emission into the clockwise (CW) and counter-clockwise (CCW) modes of a three-mirror ring cavity (red beams) \cite{Culver16}. (b) Energy level diagram. The MOT light drives lasing through Mollow gain, dominated by the $2\leftrightarrow 2^\prime$ transition \cite{SM}. Spontaneous emission here refers to free-space (non-cavity) modes.}\label{fig:schem}
\end{figure}

Our experiment is shown schematically in Fig.~\ref{fig:schem}(a). Potassium-39 atoms are cooled in a two-dimensional magneto-optical trap (MOT) and continuously transferred with a pushing laser to a three-dimensional MOT positioned at the waist of a triangular ring cavity under ultrahigh vacuum \cite{Culver16}. The trapped cloud has a typical root-mean-squared radius of $800~\mu$m and a temperature of $1~$mK. The light for cooling and pumping is near-resonant with the manifold of $|4^2{\rm S}_{1/2}, F=2\rangle \leftrightarrow |4^2{\rm P}_{3/2}, F^\prime\rangle$ transitions, and light near resonance with $F=1\leftrightarrow F^\prime$ incoherently repopulates the $F=2$ states (here $F$ is the total electronic plus nuclear angular momentum, and the prime denotes an excited state); a level diagram is shown in Fig.~\ref{fig:schem}(b). A quadrupole magnetic field is on during all measurements, with a gradient of $5.5~$G/cm in the weak direction along the cavity axis. The cavity length is $9.5$~cm and the linewidth $\kappa=2\pi\times 1.8$~MHz full-width at half-maximum (FWHM). The polarization dependence of the dielectric mirror coatings, together with the ring geometry, ensure the laser light is linearly polarized normal to the cavity plane. The emission from one side of the cavity is either imaged onto a commercial beam profiler or spatially filtered with single mode fibers and detected with a pair of analog avalanche photodiodes (APDs) or single-photon counting modules (SPCMs). In the latter case, neutral density filters are used above the lasing threshold to prevent damage to the detectors. 

Two gain mechanisms have been identified with cesium and rubidium atoms during magneto-optical trapping or under similar conditions --- Raman gain between Zeeman states within a single hyperfine level, and a nonlinear process known as Mollow gain~\cite{Guerin08, Tabosa91, Grison91, Mitsunaga96}. The latter occurs with two-level atoms driven by strong, detuned fields, and involves the absorption of two pump photons, the stimulated emission of one photon into the cavity mode, and the spontaneous emission of one photon into free space~\cite{Mollow72, Wu77} (see Fig.~\ref{fig:schem}(b)). We find that only Mollow gain exists in our MOT \cite{SM}. In potassium-39 the excited-state hyperfine splittings are on the order of the natural atomic linewidth, $\Gamma=2\pi\times 6.0$~MHz (FWHM), so that Raman gain is obscured by the nearby $2\leftrightarrow 1^\prime,3^\prime$ transitions.

\begin{figure}[ht]
\centering
\includegraphics[width=0.9\columnwidth]{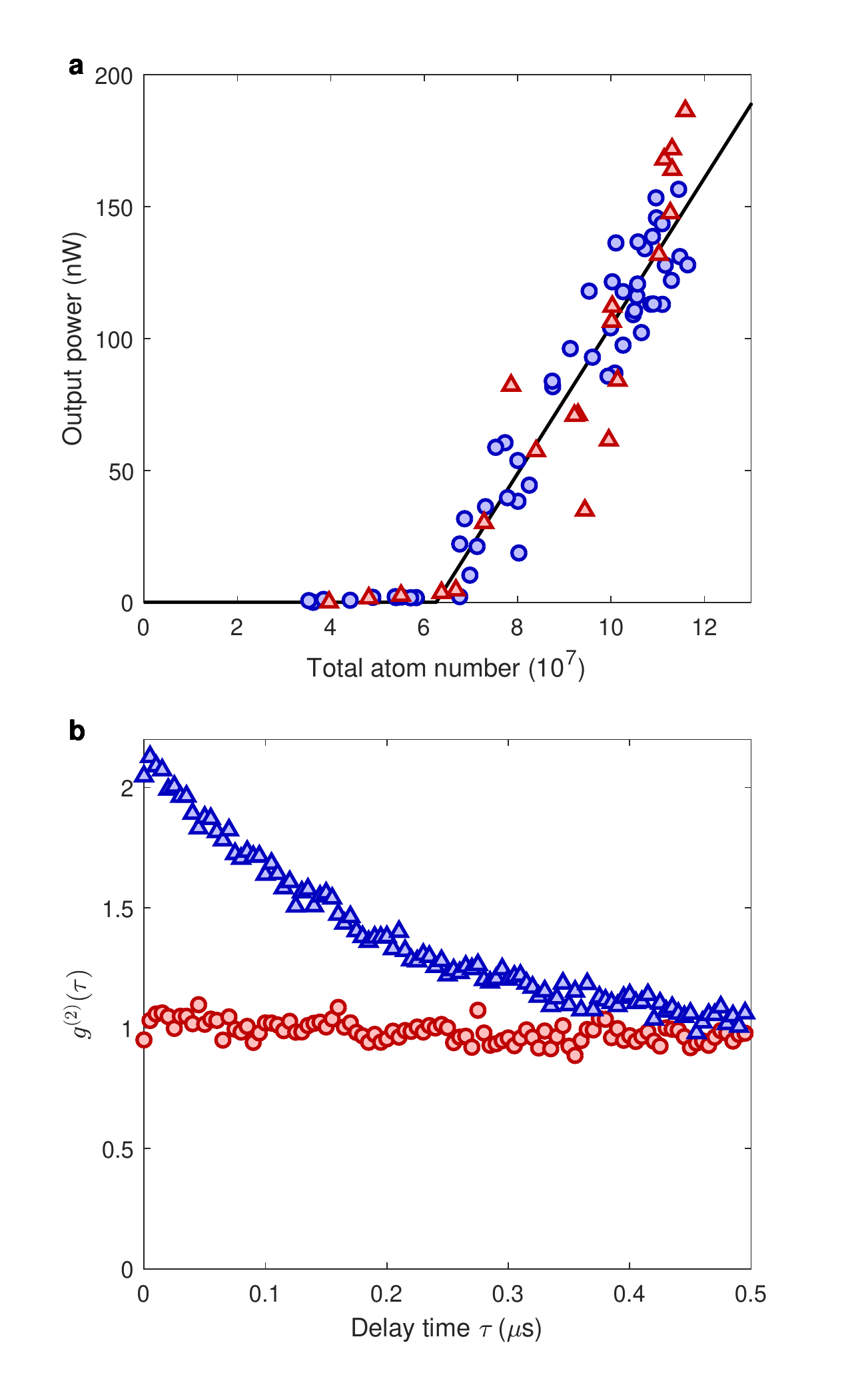}
\caption{Transition to lasing. (a) Threshold behavior of the ring laser output power as the atom number is increased, with a fixed pump intensity of $18~{\rm mW/cm}^2$  and detuning of $-27~{\rm MHz}$ from the $2\leftrightarrow 3^\prime$ transition. Blue circles (red triangles) are data for the CW (CCW) mode, and the black line is a fit to both sets of data. The onset of lasing occurs for a total atom number in the MOT of $6.2(8) \times 10^7$, and the measured slope efficiency is $2.8(2)~{\rm fW/atom}$. (b) Second-order optical coherence of the CW TEM$_{00}$ mode below ($4\times 10^7$ atoms, blue triangles) and above ($8\times 10^7$ atoms, red circles) threshold.}\label{fig:onset}
\end{figure}

Lasing can occur when the gain exceeds the fractional round-trip power loss of the cold cavity, which is below $4\times 10^{-3}$ in our experiment. As the gain depends on atom number, the lasing threshold can be crossed at constant pump intensity and detuning by varying the flux of the atomic beam loading the MOT through its dependence on the pushing laser intensity; since the cloud size is larger than the cavity waist ($90~\mu{\rm m}\times 130~\mu{\rm m}$ for the TEM$_{00}$ transverse electromagnetic modes), the effective number of atoms in the cavity is around 200 times smaller than the total number in the MOT. Typical data are shown in Fig.~\ref{fig:onset}(a). To set the scale, an output power of $0.85$~nW corresponds to the saturation intensity $I_{\rm sat}=1.75$~mW/cm$^2$ at the cavity waist (or mean number of photons of around 390). These data were obtained with the beam profiler so that all TEM modes were detected, with intensities integrated over the 1~ms exposure time. Threshold is reached first for the TEM$_{00}$ modes, after which increasing the number of atoms drives spatially multimode emission in both directions. The observed slope efficiency suggests that an atom within the cavity mode volume undergoes stimulated emission at a rate $\sim 2\times 10^6~{\rm s}^{-1}$, which is around $30\%$ of the estimated spontaneous emission rate.

To provide further evidence of lasing, the second-order optical coherence, $g^{(2)}(\tau)$, was measured for the CW TEM$_{00}$ mode above and below threshold in a Hanbury Brown-Twiss interferometer; the results are shown in Fig.~\ref{fig:onset}(b). Below threshold the photon counts exhibit super-Poissonian statistics and bunching characteristic of a thermal state, with $g^{(2)}(0)=2.160(15)$ and a coherence time of $170(5)$~ns. This time scale is twice the measured energy decay time of the cavity ($1/\kappa$). The fact that the coherence time is longer than $1/\kappa$ can be understood as a consequence of the below-threshold gain medium, which partially compensates the round-trip losses in the cavity \cite{SM}. Above threshold, we find $g^{(2)}(\tau)\simeq 1$, as expected for an ideal laser, provided we account for the random directional switching described in detail below \cite{SM}. 

A homogeneously broadened ring laser cannot emit simultaneously into both directions, due to competition for gain in the saturated medium~\cite{SargentScullyLamb,Singh79}. The dimensionless third-order equations of motion for the lasing fields are,
\begin{eqnarray}\nonumber
\dot{E}_1 &=& \left[a_1+p_1(t)-|E_1|^2-\xi|E_2|^2\right]E_1+q_1(t) \quad, \\
\dot{E}_2 &=& \left[a_2+p_2(t)-|E_2|^2-\xi|E_1|^2\right]E_2+q_2(t) \quad.
\label{eq:Lamb}
\end{eqnarray}
Here 1 and 2 label the CW and CCW modes, respectively, and the $a_i$ are the corresponding pump parameters, equal to the ratio of pump intensity to threshold pump intensity minus 1. The $p_i$ and $q_i$ are Langevin noise terms describing the effects of pump fluctuations and spontaneous emission, respectively. For a homogeneously broadened laser, the cross-coupling constant $\xi$ can be as large as 2, leading to the possibility of bistable behavior.

\begin{figure}[ht]
\centering
\includegraphics[width=\columnwidth]{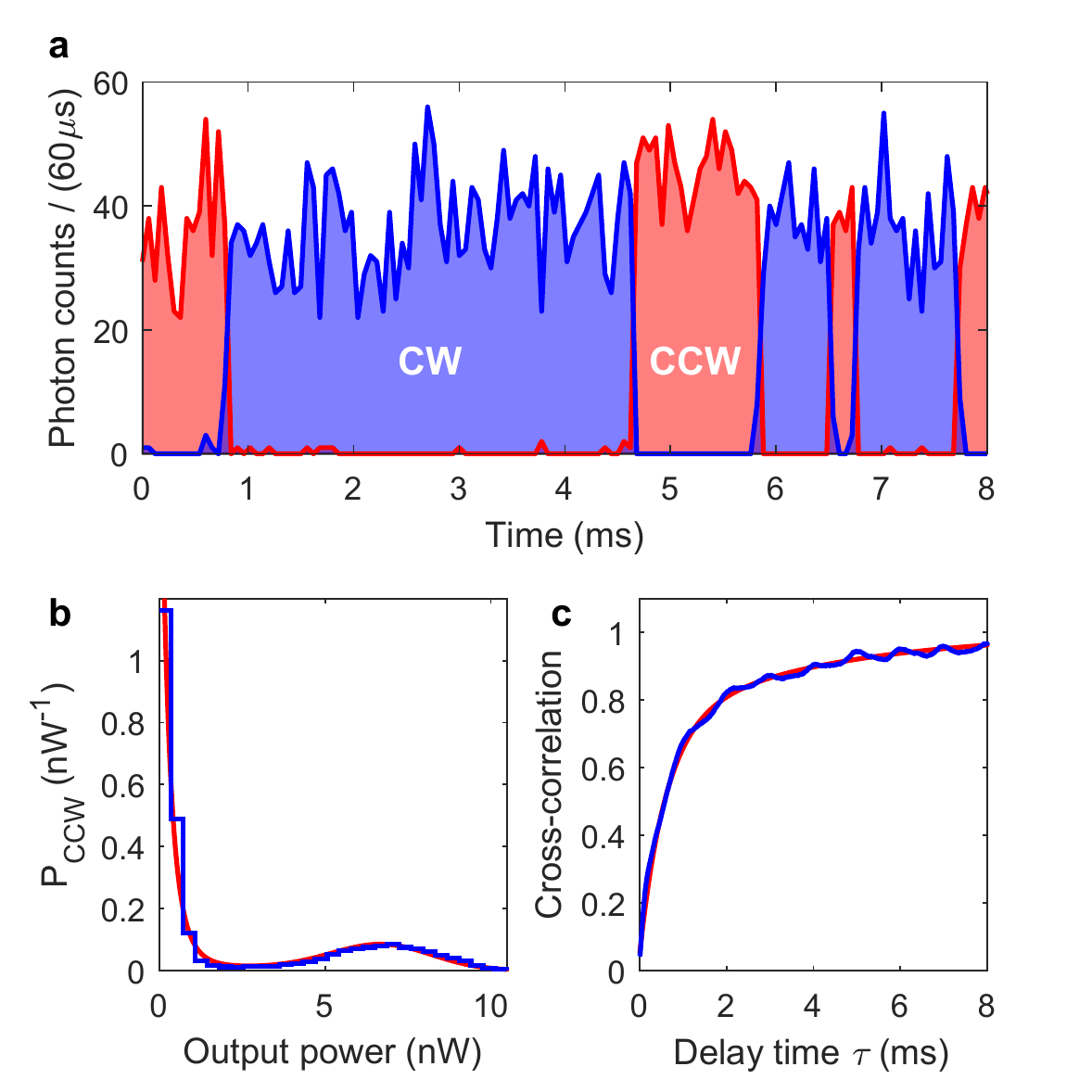}
\caption{Directional bistability. (a) Random switching between modes. Photon counts from each direction, integrated over $60~\mu{\rm s}$, revealing quasi-continuous unidirectional lasing. (b) Probability density $P_{\rm CCW}$ for lasing into the CCW direction, obtained from the analog APD signal without attenuation. Data are in blue and the theoretical prediction from Eqs.(\ref{eq:Lamb}) is in red. (c) Normalized cross-correlation between directions, obtained from the photon counts with a resolution of $10~\mu{\rm s}$. Zero corresponds to perfect anti-correlations and 1 to uncorrelated counts. Data are in blue and the red curve shows a double exponential with time constants $0.6$ and $4$~ms.}\label{fig:bistab}
\end{figure}

We observe such directional bistability in the cold atom ring laser, as shown in Fig.~\ref{fig:bistab}(a). The laser emits quasi-continuously into one mode at a time, switching between CW and CCW directions at random intervals. As discussed in earlier work on dye ring lasers, this switching is dominated by the quantum noise terms $q_i(t)$, even when the pump noise terms $p_i(t)$ are relatively large~\cite{Lett81,Lett85}. As such, the switching represents a remarkably macroscopic effect of individual spontaneous emission events. In Figure \ref{fig:bistab}(b) we show the probability density for the CCW lasing power obtained from one of the APDs. The bimodal distribution is clear, reflecting the nearly perfect on/off nature of the switching. The experimentally determined probability density is in excellent agreement with the theoretical prediction obtained from Eqs.(\ref{eq:Lamb}) \cite{Singh79}. The calculation assumes a mean pump parameter $a=(a_1+a_2)/2=8.7$ and difference $\Delta a=a_1-a_2=0.24$ (we show below how the asymmetry $\Delta a$ can be controlled in our experiment). The normalized cross-correlation between CW and CCW directions, shown in Fig.~\ref{fig:bistab}(c), highlights the strong suppression of simultaneous emission, being only $0.04$ at zero delay time. The correlation tends to 1 on a time scale of order one millisecond, reflecting the characteristic dwell time for continuous emission into either direction. Specifically, the data are well described by a double exponential curve with time constants of $0.6$ and $4$~ms. The existence of two such time scales in a semiconductor ring laser was interpreted in terms of the phase-space topology of the solutions to Eqs.(\ref{eq:Lamb}) in \cite{Beri08}.

\begin{figure*}[ht]
\centering
\includegraphics[width=0.8\textwidth]{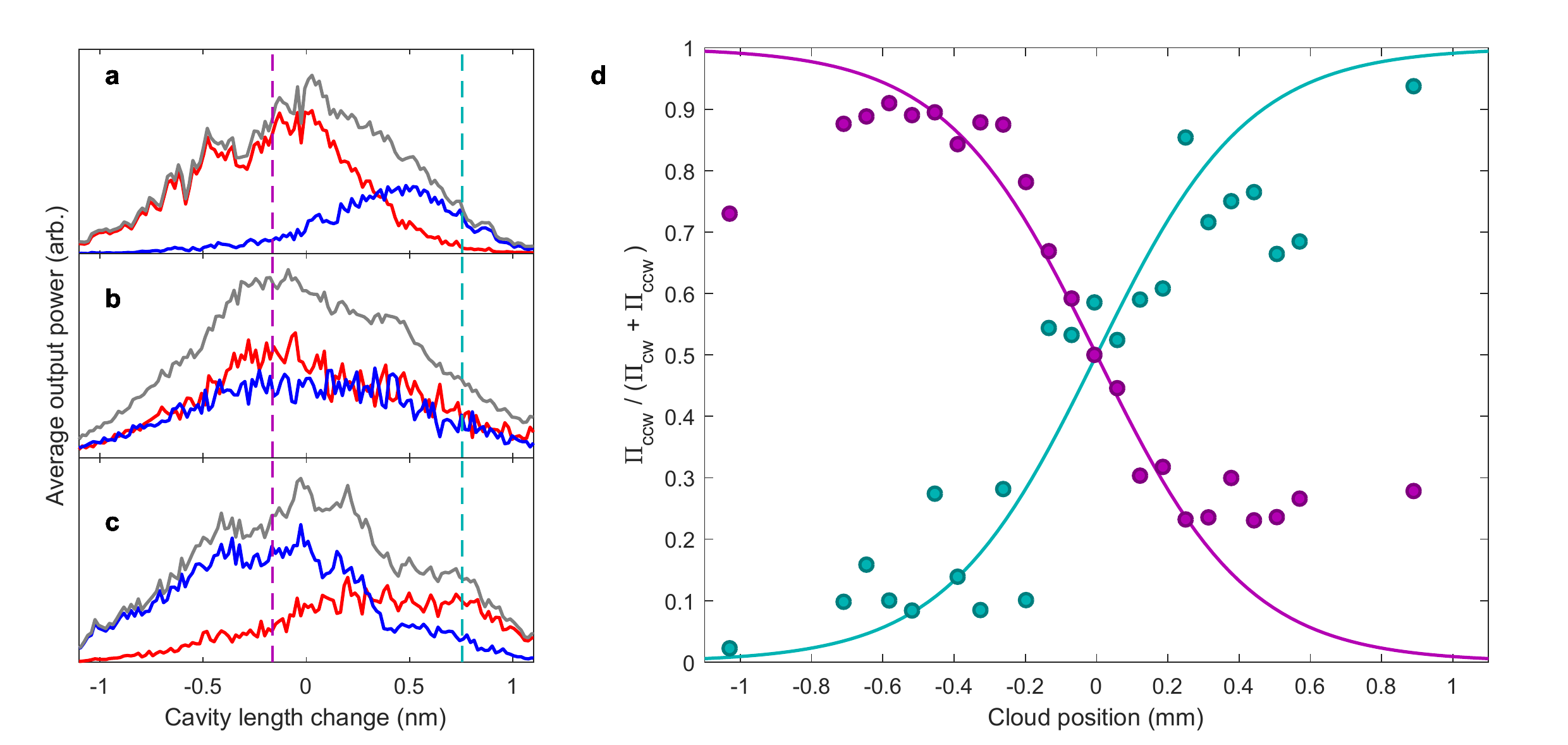}
\caption{Nonreciprocal lasing. (a)--(c) Time-averaged output power of the CW mode (blue), CCW mode (red), and sum (gray), as the cavity length is scanned for different cloud positions. The position in (a), (b), and (c) was $-0.51$, $0.00$, and $0.51$~mm, respectively, referenced to the observed center of symmetry. (d) Normalized lasing probability for the CCW direction. Points are data and lines are the theoretical predictions derived from Eqs.(\ref{eq:Lamb}); in both cases the color magenta or cyan corresponds to the cavity lengths highlighted with vertical dashed lines in (a)--(c).}\label{fig:nonrecip}
\end{figure*}

The non-zero value of $\Delta a$ inferred from Fig.~\ref{fig:bistab}(b) implies an asymmetry between the counter-propagating TEM$_{00}$ modes. In dye ring lasers, nonreciprocity has been controlled by incorporating an intracavity acousto-optic modulator~\cite{McNamara88} or Faraday rotator~\cite{Gage88}. In our laser, the directional asymmetry can be tuned by moving the cloud. This is done by adding a uniform magnetic field along the cavity axis and varying its magnitude. In figures~\ref{fig:nonrecip}(a)--(c), we show the time-averaged output powers as the cavity length is scanned for three different cloud positions. As the cloud moves, the optimum cavity lengths for the CW and CCW modes shift relative to one another, while the summed power remains stationary. The changes in average power are dominated by changes in the average duration of lasing, with relatively little variation in peak pulse power. The amplitude of the right peak (i.e., the peak with longer optimum cavity length) is always smaller than the left due to the influence of nearby higher-order TEM modes which are present but suppressed in the detected signal by the single mode fibers. The maximum observed shift between optimum cavity lengths corresponds to a frequency nonreciprocity of $\sim 2$~MHz when referenced to the empty cavity tuning, which is much larger than what is observed in conventional ring lasers~\cite{Kravtsov99}. Figure \ref{fig:nonrecip}(d) shows how the CCW lasing probability $\Pi_{\rm CCW}$ varies with cloud position for two different values of cavity length. For comparison, the solid curves show the theoretical prediction derived from Eqs.(\ref{eq:Lamb}) \cite{Lett81} under the empirically motivated assumption that $\Delta a$ changes linearly with cloud position while $a$ remains constant. 

The lasing asymmetry varies with position over a length scale which is comparable to the cloud size (the Rayleigh range of the cavity modes is around $50\times$ longer). When the magnetic field gradient is changed, the symmetry point also moves, suggesting that the location of reciprocal lasing is not dictated by the geometry of the cavity modes. Naively one expects the trapped cloud to be located at the zero of the magnetic field, but experience shows that the cloud position also depends in a non-trivial way on the alignment, intensities, and polarizations of the cooling laser beams. We believe that the true position reflects a compromise between the location where $B=0$ and the complex influence of the radiation pressure, and that reciprocal lasing occurs when the cloud is exactly centered on the magnetic field. When the cloud is displaced from this position, nonreciprocity could arise due to the varying strength and direction of the magnetic field across the cloud and/or an asymmetry among the Zeeman state populations from optical pumping. Such imbalances can lead to Faraday rotation and nonreciprocal losses such as those harnessed in Zeeman laser gyros with uniform applied fields \cite{Sargent67}.

We can now compare our observations with previous experiments. To our knowledge only two experiments have reported lasing with MOTs operating under normal conditions and without additional fields, in both cases using standing-wave cavities. The first cold-atom laser employed cesium atoms with a low-finesse cavity built around the vacuum chamber~\cite{Hilico92}. Heterodyne measurements of the emission identified Raman gain as the underlying mechanism. As mentioned above, we do not observe Raman gain in potassium-39, as the pump light is always near-resonant with multiple transitions. Recently a rubidium-87 MOT was made to lase in the collective strong coupling regime of cavity QED~\cite{Sawant17}. The gain was attributed to a Mollow-type mechanism driven by the combined pump and cavity fields. Although broadly similar to our experiment in terms of cavity QED parameters, our results are qualitatively different in many aspects. Most significantly, we do not observe doublets in the emission as the cavity length is scanned, and our lasing power is several orders of magnitude brighter than the Purcell-enhanced scattering below threshold.

In the future we aim to achieve simultaneous bidirectional lasing either by inducing inhomogeneous broadening or by pumping the two directions separately using four-wave mixing~\cite{Guerin08}. This would open up the possibility of active rotation sensing with cold atoms. Finally, we plan to investigate what role, if any, light-matter coherence can play in our system. The $N$-atom vacuum Rabi frequency far exceeds the decay rates $\kappa$ and $\Gamma$, as well as the excited-state hyperfine and TEM mode splittings \cite{Culver16}. We have observed that in the absence of lasing the vacuum Rabi splitting survives the dissipative processes acting within the MOT, motivating a search for evidence of Rabi oscillations or coherent inter-mode coupling in the laser emission.

\begin{acknowledgments}
B.M.~and G.H.~contributed equally to this work. The apparatus was built with funding from the UK EPSRC (EP/J016985/1), and B.M.~is supported through DSTL (DSTLX1000092132). We are grateful to Vincent Boyer, Giovanni Barontini, and Kai Bongs for loaning essential hardware and for useful discussions, and to Yu-Hung Lien for feedback on the manuscript. Robert Culver and Sam Goldwin assisted with making Fig.~\ref{fig:schem}(a). The data supporting this work are available from the corresponding author upon reasonable request.
\end{acknowledgments}


\pagebreak
\clearpage
\begin{center}
\textbf{\large Supplemental Material}
\end{center}

\setcounter{equation}{0}
\setcounter{figure}{0}
\setcounter{table}{0}
\setcounter{page}{1}
\makeatletter
\renewcommand{\thefigure}{S\arabic{figure}}
\renewcommand{\thesection}{S\arabic{section}}
\renewcommand{\theequation}{S\arabic{equation}}
\renewcommand{\thepage}{S\arabic{page}}
\renewcommand{\bibnumfmt}[1]{[S#1]}
\renewcommand{\citenumfont}[1]{S#1}

~\\
Here we describe transmission measurements and theoretical modeling which support the conclusion that lasing is sustained by Mollow gain. We then present our analysis of the effect of bistable lasing on the second-order coherence, and a toy model describing the extended coherence time of bunching below threshold.
~\\

\section{Gain profile\label{sec:gain}}

\begin{figure*}[ht]
\centering
\includegraphics[width=\textwidth]{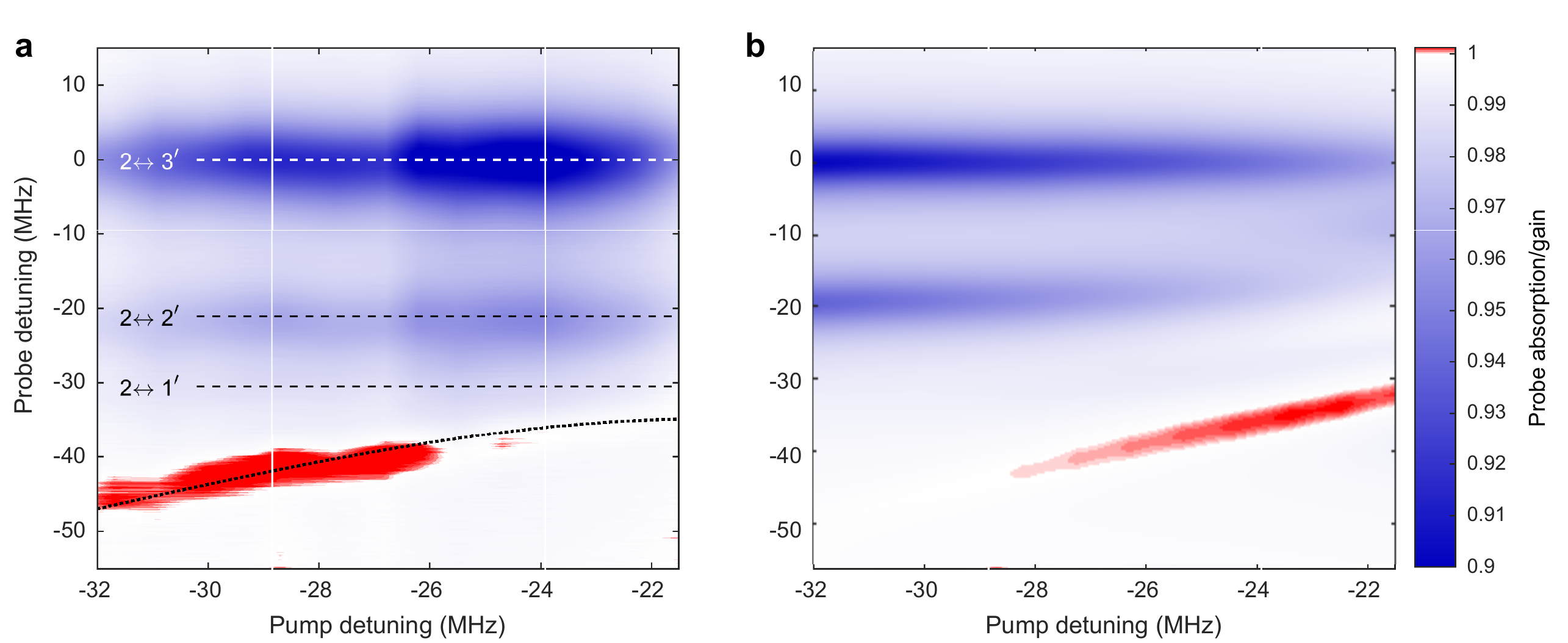}
\caption{Gain spectrum. Data are shown in (a) and theory in (b). Blue corresponds to absorption and red to gain. In (a) the dashed lines show the calculated transition frequencies in the absence of Stark shifts from the pump light. The dotted curve shows the expectation from Eq.(\ref{eq:omegapm}). The calculated spectrum in (b) is obtained from equations (\ref{eq:Hprime})--(\ref{eq:diss}) for the four-level model. The parameters were chosen to match the experimental conditions. The simulated spectrum was scaled with an overall multiplicative factor to match the peak absorption from the data, and the probe detunings were referenced to the Stark-shifted $2\leftrightarrow 3^\prime$ resonance.}\label{fig:gain}
\end{figure*}

In order to measure the gain profile, the trapped cloud was displaced from the cavity by applying a uniform magnetic field transverse to the cavity axis. A probe beam of a few microwatts was focused to a waist of $400~\mu$m at the cloud and the transmission was measured with an avalanche photodiode. For each value of pump detuning, the probe laser frequency was scanned and the average of 64 spectra was recorded on a digital oscilloscope. This was repeated for pump detunings incremented in steps of 1--2~MHz. The dominant $2\to 3^\prime$ absorption features were used to align the individual spectra, as the center frequency of the probe laser was not stabilized during the measurements. The results are shown in Fig.~\ref{fig:gain}(a). Raman gain would occur for probe frequencies near the pump frequency; as no such features are observed, we rule out Raman gain in this system.

Gain in a strongly driven two-level atom was described theoretically by Mollow in \cite{Mollow72SM} and observed by Wu et al.~in \cite{Wu77SM}. The absorption coefficient is proportional to,
\begin{eqnarray}\label{eq:Mollow2LA}
\alpha &=& \frac{\Gamma}{2}\frac{|z|^2}{|z|^2+\Omega^2/2}\\ \nonumber
&\times& {\rm Re}\! \left[ \frac{(\Gamma+i\delta)(z+i\delta)-i\Omega^2\delta/(2z)}{(\Gamma+i\delta)(z+i\delta)(z^\ast+i\delta) + \Omega^2(\Gamma/2+i\delta)} \right]\;,
\end{eqnarray}
where $\Delta<0$ is the pump detuning, $z=\Gamma/2-i\Delta$, $\delta$ is the probe detuning, and all other parameters are as defined in the main text (this formula also appears in \cite{Guerin08SM}). For strong, off-resonant driving, the spectrum is dominated by a pair of gain and absorption features lying symmetrically around the pump frequency $\omega_p$,
\begin{eqnarray}
\omega_\pm \simeq \omega_p \pm \Delta\sqrt{1+(\Omega/\Delta)^2}\quad,
\label{eq:omegapm}
\end{eqnarray}
where $\Omega=C\,\Gamma\sqrt{I_p/(2I_{\rm sat})}$ is the Rabi frequency for a pump intensity $I_p$ and $C$ is the Clebsch-Gordan coefficient. As shown in Figure \ref{fig:gain}(a), this simple formula does a reasonable job predicting the observed center frequency of the gain peak if only $2\leftrightarrow 2^\prime$ scattering is considered. The gain amplitude predicted by Eq.(\ref{eq:Mollow2LA}), however, is far smaller than what is observed in the experiment. If one adds to this expression for $\alpha$ the two-level expressions for absorption from $2\leftrightarrow 1^\prime$ and $3^\prime$, the result fails to predict net gain at any frequency within our parameter range.

Better agreement with the observed spectrum is found if we repeat the derivation leading to Eq.(\ref{eq:Mollow2LA}), starting with a multi-level Hamiltonian. Rather than considering all possible Zeeman states, we adopt a four-level model in the basis $\{|2\rangle,|1^\prime\rangle,|2^\prime\rangle,|3^\prime\rangle\}$, and scale the Rabi frequencies by the Clebsch-Gordan coefficents for $\pi$ transitions averaged over all Zeeman states, which we denote $C_{FF^\prime}$. Specifically, $(C_{21^\prime},C_{22^\prime},C_{23^\prime})=(\sqrt{1/30},\sqrt{1/6},\sqrt{7/15})$. For example, the Hamiltonian for a weak probe field with frequency $\omega^\prime$ is then,
\begin{eqnarray}
\hat{H}^\prime = \frac{\hbar\Omega^\prime}{2}\left(\hat{\Sigma}e^{i\omega^\prime t} + \hat{\Sigma}^\dagger e^{-i\omega^\prime t}\right)\quad,
\label{eq:Hprime}
\end{eqnarray}
where $\hat{\Sigma} = \sum_{F^\prime}C_{2F^\prime}|2\rangle\langle F^\prime|$ is the weighted sum of atomic lowering operators. Working in the interaction picture and rotating wave approximation, the absorbed power is proportional to,
\begin{eqnarray}\nonumber
\mathcal{P} &=& \left\langle\partial_t\hat{H}^\prime(t)\right\rangle \\
&=& \hbar\omega^\prime\left(\frac{\Omega^{\prime}}{2}\right)^{\!\!2}\int_{-\infty}^\infty d\tau\left\langle [\hat{\Sigma}(\tau), \hat{\Sigma}^\dagger(0)]\right\rangle e^{i\omega^\prime\tau}\;.
\label{eq:spectrum}
\end{eqnarray}
The average $\langle\cdot\rangle$ is obtained from the master equation for the density operator $\rho$, with respect to the pump Hamiltonian $\hat{H}_0$ in the absence of the probe ($\hat{H}_0$ has the same form as $\hat{H}^\prime$, but with Rabi and optical frequencies associated with the pump light),
\begin{eqnarray}
\dot{\rho}(t) = \frac{1}{i\hbar}[\hat{H}_0(t),\rho(t)] + \Gamma\sum_{F^\prime}\mathcal{D}\left[\,|2\rangle\langle F^\prime|\right]\rho(t),
\label{eq:master}
\end{eqnarray} 
where the dissipation superoperator $\mathcal{D}[\hat{A}]\rho$ for a collapse operator $\hat{A}$ is,
\begin{eqnarray}
\mathcal{D}[\hat{A}]\rho = \hat{A}\rho\hat{A}^\dagger-\frac{1}{2}\left(\rho\hat{A}^\dagger\hat{A} + \hat{A}^\dagger\hat{A}\rho\right)\quad.
\label{eq:diss}
\end{eqnarray}

Equations (\ref{eq:Hprime})--(\ref{eq:diss}) were evaluated numerically using the QuTiP software package \cite{Johansson12SM, Johansson13SM}. The results are shown in Fig.~\ref{fig:gain}(b). The simulation accurately captures both the existence of net gain and its center frequency. Although the simulation shows increasing gain for smaller pump detunings, in contrast to what is observed in the experiment, the calculation does not account for the variation in atom number in the MOT as the pump detuning changes.

\section{Second-order coherence} 

The second-order coherence of the CW output was measured in a Hanbury Brown-Twiss interferometer comprising a non-polarizing 50:50 beam splitter and a pair of SPCMs. Time stamps from a total of $10^6$--$10^7$ photon counts were recorded with $81$~ps resolution, and a histogram of the time differences between successive counts at different detectors was produced with $5$~ns resolution. The value of $g^{(2)}(\tau)$ was obtained by normalizing the binned coincidence counts to the product of the mean counts in each channel. As mentioned in the main text, neutral density filters were used to limit the count rate above threshold, which does not affect the measured values of $g^{(2)}$.

\subsection{Effect of bistability}

If we evaluate $g^{(2)}$ above threshold for count streams of tens of milliseconds, we find $g^{(2)}(\tau)\sim 5$, dominated by the bistable switching shown in Fig.~3(a) of the text and relatively slow fluctuations in the ring laser output power. The data above threshold were therefore corrected as follows. Over short times the laser emission is assumed to be in a pure coherent state, subject to directional switching and relatively slow fluctuations in average output power. This results in a conditional probability distribution for the photon number $n$ given by $p(n|RT)=(RT)^n\exp(-RT)/n!$, where $R$ is the randomly varying instantaneous count rate and $T$ is a time which is much greater than the range of $\tau$ considered, but shorter than the time scale of the switching. Using this conditional distribution, the probability distribution for $n$ is $p(n)=\sum_R p(R)\,p(n|RT)$, with $p(R)$ the probability density for $R$, from which one finds,
\begin{eqnarray*}
\frac{{\rm var}(n)}{\langle n\rangle} &=& 1+T\frac{{\rm var}(R)}{\langle R\rangle} \quad.
\end{eqnarray*}
Using the fact that,
\begin{eqnarray*}
g^{(2)}(0) = 1+\frac{1}{\langle n\rangle}\left[ \frac{{\rm var}(n)}{\langle n\rangle} - 1\right] \quad,
\end{eqnarray*}
and choosing $T=\langle n\rangle/\langle R\rangle$, we obtain a correction factor $\chi=1+{\rm var}(R)/\langle R\rangle^2$. Three series of time stamps were recorded under similar conditions above threshold, and $[g^{(2)}(\tau)/\chi]$ was averaged over the set at each $\tau$ to obtain the points in Fig.~2(b) of the text. We stress that this correction only rescales the magnitude of $g^{(2)}(\tau)$; the lack of dependence on $\tau$ is already a feature of the raw histograms.~\\

\subsection{Coherence time of bunching}

\begin{figure}[ht]
\centering
\includegraphics[width=\columnwidth]{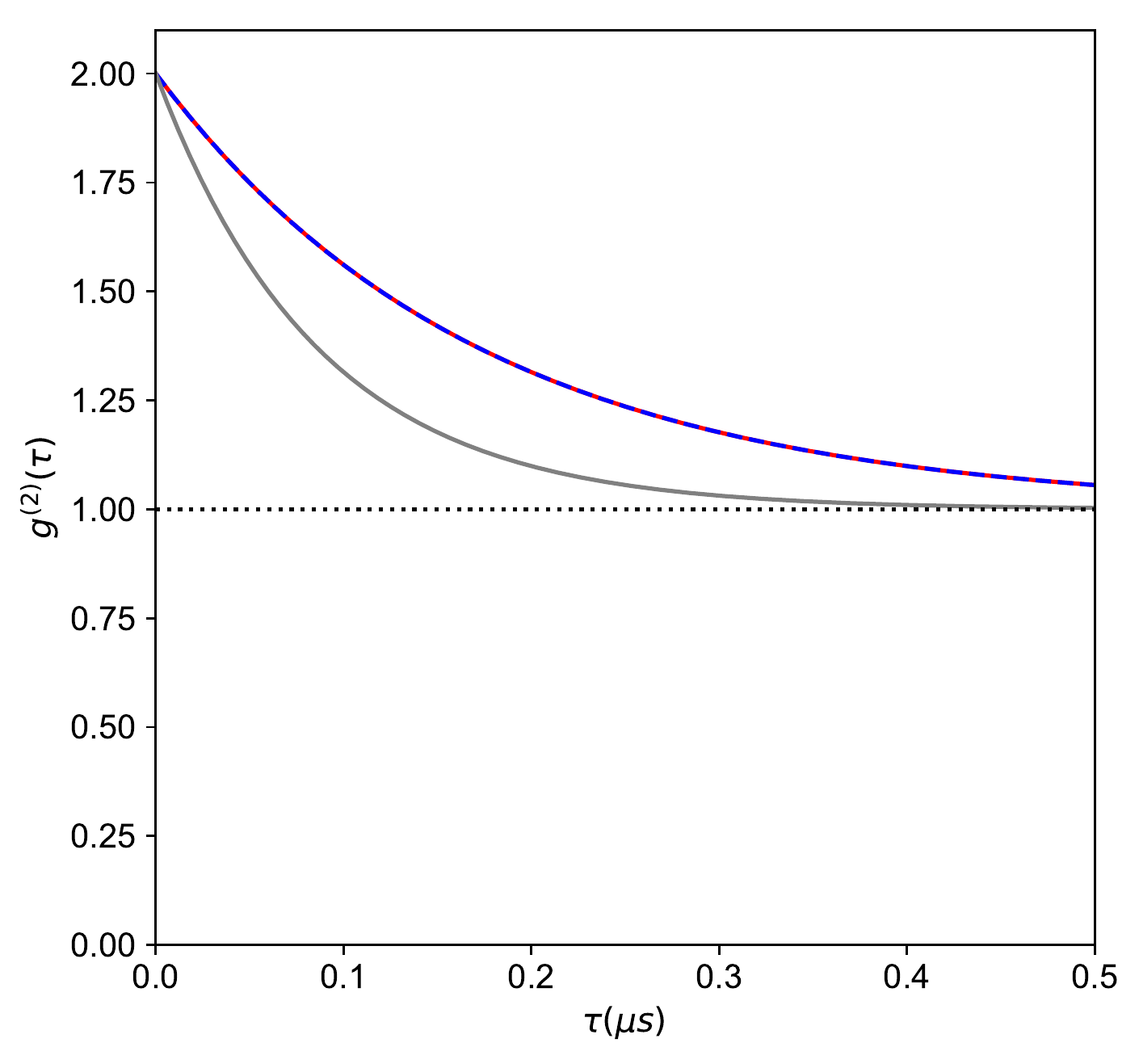}
\caption{Second-order coherence time. The solid red curve shows $g^{(2)}(\tau)$ obtained numerically for gain reservoir coupling with strength $G=\kappa/2$ and $\bar{n}=0$, and the overlapping dashed blue curve shows $1+\exp(-\kappa\tau/2)$. For comparison, the solid gray curve shows a decay constant $1/\kappa$ and the dotted line shows $g^{(2)}(\tau)=1$, characteristic of coherent pumping of the cavity field with arbitrary rate (described by replacing the reservoir coupling $G\mathcal{D}[\hat{a}^\dagger]\rho$ with a term $\hbar G(\hat{a}+\hat{a}^\dagger)$ in the Hamiltonian).}\label{fig:coherence}
\end{figure}

In Fig.~2(b) of the main text we show that the second-order coherence time below threshold exhibits bunching with a decay time constant approximately equal to $2/\kappa$, where $1/\kappa$ is the energy decay time of the cold cavity. In our experiment, with $\kappa < \Gamma$, we expect the cavity to smooth out fluctuations in the stimulated emission, but it still may be surprising that the coherence time could exceed $1/\kappa$. This behavior can be understood by assuming the gain medium acts as an incoherent pump for the cavity field. As described in \cite{QOSM}, the two-level gain medium can be modeled as a bath of inverted harmonic oscillators supplying the energy needed for amplification. Specifically, we take the Hamiltonian for the free cavity field, $\hbar\omega_c\,\hat{a}^\dagger\hat{a}$, where $\omega_c$ is the frequency and $\hat{a}$ is the usual photon annihilation operator. We then solve the master equation with cavity decay described by $\kappa\mathcal{D}[\hat{a}]\rho$ and coupling to the gain reservoir by $G(\bar{n}+1)\mathcal{D}[\hat{a}^\dagger]\rho+G\bar{n}\mathcal{D}[\hat{a}]\rho$, with $G$ the coupling strength and $\bar{n}$ the mean number of reservoir photons. Assuming the amplifier is in its vacuum state ($\bar{n}=0$) leaves only the spontaneous emission noise term. This gives a second-order coherence characteristic of a thermal field with bunching coherence time equal to $1/(\kappa-G)$; we have verified experimentally that the coherence time increases with gain. An example calculation is shown in Fig.~\ref{fig:coherence}. Note that this model predicts $g^{(2)}(0)=2$, as observed in the experiment. Although this model lacks microscopic detail with respect to the driven atoms, the interpretation is clear --- the gain extends the coherence time by mitigating losses, while spontaneous emission noise within the amplifier leads to bunching.

\end{document}